\documentclass[aps,prl,twocolumn,groupedaddress]{revtex4-1}
\usepackage{amsmath}
\usepackage{graphicx}
\DeclareGraphicsExtensions{.pdf,.eps,.png,.jpg,.mps}
\usepackage{epstopdf}
\usepackage{threeparttable}
\usepackage{color}

\DeclareRobustCommand{\rchi}{{\mathpalette\irchi\relax}}
\newcommand{\irchi}[2]{\raisebox{\depth}{$#1\chi$}}

\begin{document}
\title{Hot electrons modulation of third harmonic generation in graphene}
\author{G. Soavi$^1$}
\thanks{Present Address: Institut fur Festkorperphysik, Friedrich-Schiller-Universitat, Max-Wien-Platz 1, 07743 Jena}
\author{G. Wang$^1$, H. Rostami$^2$, A. Tomadin$^3$, O. Balci$^1$, I. Paradisanos$^1$, E.A.A. Pogna$^4$, G. Cerullo$^4$, E. Lidorikis$^5$, M. Polini$^3$, A. C. Ferrari$^1$}\email{acf26@eng.cam.ac.uk}
\affiliation{$^1$ Cambridge Graphene Centre, University of Cambridge, Cambridge CB3 0FA, UK}
\affiliation{$^2$ Nordic Institute for Theoretical Physics, Roslagstullsbacken 23 SE-106 91 Stockholm, Sweden}
\affiliation{$^3$ Istituto Italiano di Tecnologia, Graphene Labs, Via Morego 30, I-16163 Genova, Italy}
\affiliation{$^4$IFN-CNR, Dipartimento di Fisica, Politecnico di Milano, P.zza L. da Vinci 32, 20133 Milano, Italy}
\affiliation{$^5$ Department of Materials Science and Engineering, University of Ioannina, Ioannina 45110, Greece}
\begin{abstract}
Hot electrons dominate the ultrafast ($\sim$fs-ps) optical and electronic properties of metals and semiconductors and they are exploited in a variety of applications including photovoltaics and photodetection. We perform power-dependent third harmonic generation measurements on gated single-layer graphene and detect a significant deviation from the cubic power-law expected for a third harmonic generation process. We assign this to the presence of hot electrons. Our results indicate that the performance of nonlinear photonic devices based on graphene, such as optical modulators and frequency converters, can be affected by changes in the electronic temperature, which might occur due to increase of absorbed optical power or Joule heating.
\end{abstract}
\maketitle
For a free electron gas at thermal equilibrium, the average occupation number at energy $E$ is described by the Fermi-Dirac distribution $f(E)$\cite{kittel1996}:
\begin{equation}
\label{eq:Fermi-Dirac}
f(E)=\frac{1}{e^{(E-\mu)/k_BT_0}+1}~,
\end{equation}
where $\mu$ is the chemical potential and $k_B$ is the Boltzmann constant. At zero temperature, $\mu$ equals the Fermi energy ($E_F$). At thermal equilibrium $T_e=T_l=T_0$, with $T_e$ the electronic temperature, $T_l$ the lattice temperature and $T_0$ the ambient temperature. Photoexcitation of a sample with ultrashort ($\sim$fs-ps) pulses creates a non-thermal regime, i.e. a condition where the electron population cannot be defined by $f(E)$ and $T_e$, which rapidly evolves through electron-electron (e-e) scattering into a hot-carrier distribution, with $T_e>T_l$\cite{farm1992,dellavalle2012,lazzeriPRL2005,bridaNC2013,tomadinPRB2013}. Electrons then transfer energy to the lattice through scattering with phonons (ph) until $T_e=T_l$\cite{lazzeriPRL2005,bridaNC2013,tomadinPRB2013,shank1983,schoenlein1987}. Equilibrium with the surrounding environment is then reached via ph-ph scattering\cite{lazzeriPRL2005,bridaNC2013,tomadinPRB2013,vallee2001,hohlChemPhys2000,shah2013,soaviAOM2016,weiNanop2017}. The timescale of these scattering processes depends on the system under investigation and the excitation energy. Typical values for metals (e.g. Au, Ag, Cu, Ni\cite{dellavalle2012,vallee2001,hohlChemPhys2000,shah2013}) and semiconductors (e.g. Si\cite{shank1983}) are$\sim$10fs-1ps for e-e scattering\cite{dellavalle2012},$\sim$1-100ps for e-ph scattering\cite{shank1983,schoenlein1987}, and$>$100ps for ph-ph scattering\cite{vallee2001,hohlChemPhys2000,shah2013}.

Hot electrons (HEs) can be exploited to enhance the efficiency of photocatalysis\cite{mukhNanoLett2013}, photovoltaic devices\cite{rossJAP1982,tisdScience2010} and photodetectors\cite{weiNanop2017}. The efficiency of photovoltaic devices can be enhanced if HEs are collected before relaxation with ph\cite{tisdScience2010}, when the absorbed light energy is transferred to the lattice instead of being converted into an electrical signal. Photodetectors based on the Seebeck effect\cite{stiensProcSPIE2006} and Schottky junctions\cite{shephProcIEEE1970} both exploit HEs. These also play a key role in nonlinear effects, e.g. in Second Harmonic Generation (SHG)\cite{frankPRL1961} and in Third Harmonic Generation (THG)\cite{terhPRL1962}. Following interaction with photons with energy $\hbar\omega_0$, where $\hbar$ is the reduced Planck constant and $\omega_0$ is the photon angular frequency, new photons can be generated inside a nonlinear material at energies $2\hbar\omega_0$ for SHG\cite{frankPRL1961} or $3\hbar\omega_0$ for THG\cite{terhPRL1962}. In the scalar form, the SHG and THG optical electric field ${\cal E}_{m \omega_{0}}$ can be written as\cite{shen1984,boyd2003}:
\begin{equation}
\label{eq:SHG_THG}
{\cal E}_{m\omega_0}=g \rchi^{(m)}(\omega_0,E_F,T_e) {\cal E}_{\omega_0}^m~,
\end{equation}
where ${\cal E}_{\omega_0}$ is the incident electric field, $m=2$ for SHG and $m=3$ for THG, $g$ is a function of the material's refractive index ($n$) and $\omega_0$, and $\rchi^{(m)}$ is the material's nonlinear susceptibility. $g$ and $\rchi^{(m)}$ depend on material, angle and polarization of the incident light and on m\cite{shen1984,boyd2003,kumarPRB2013}. E.g., the THG field for a bulk sample for normal incidence and constant incident power is\cite{shen1984,boyd2003,kumarPRB2013}:
\begin{equation}
\label{eq:THG_bulk}
{\cal E}_{3\omega_0}=\frac{1}{4} \frac{i3\omega_0}{2n_{3\omega_0}c}d\rchi^{(3)} {\cal E}_{\omega_0}^3~,
\end{equation}
where $d$ is the material's thickness. The light intensity ($I_{m\omega_0}$ in units of $W/m^2$) is related to the optical electric field by  $I_{m\omega_0}=n_{m\omega_0}\epsilon_0c|{\cal E}_{m\omega_0}|^2/2$\cite{shen1984,boyd2003,kumarPRB2013}. Eq.(\ref{eq:SHG_THG}) highlights two aspects of harmonic generation: (i) the SHG/THG electric field scales with the square/cube of ${\cal E}_{\omega_0}$ and, as a consequence, one would expect $I_{m\omega_0} \propto I_{\omega_0}^m$; (ii) SHG/THG intensities depend on the linear (e.g. absorption) and nonlinear (through the nonlinear susceptibilities $\rchi^{(2)}$ and $\rchi^{(3)}$) properties of the material\cite{hohlApplPhysA1995,burnsPRB1971}. Both $g$ and $\rchi^{(m)}$ are functions of $T_e$, and thus modify the power-law relation between $I_{m\omega_0}$ and $I_{\omega_0}^m$. The role of HEs in nonlinear optics was investigated for SHG in metals\cite{hohlChemPhys2000,hohlApplPhysA1995,guoPRL2001,hohlAPB1996,mooreOptLett1999,papaOptComm1997} and semiconductors\cite{tomPRL1988,saePRL1991} but, to the best of our knowledge, has not been considered thus far for THG in any material.

HEs play also a key role in the ultrafast (fs-ps)\cite{bistritzerPRL2009,betzPRL2012,bridaNC2013,tomadinPRB2013,tielrooijNP2013} and nonlinear\cite{mikhailovPRB2016,rostamiPRB2016,chengNJP2014} properties of single-layer graphene (SLG). In SLG e-e scattering occurs within few tens of fs after photoexcitation\cite{bridaNC2013}, while e-ph scattering takes place on a $\sim$ps timescale\cite{bridaNC2013,tielrooijNP2013,lazzeriPRB2006}. HEs can be exploited for the development of optoelectronic devices based on graphene\cite{bonaccorsoNP2010,RomaNRM3,FerrN2015}. E.g., a SLG p-n junction can be used as a photothermal detector because, following optical excitation, the photo-thermoelectric (or Seebeck) effect (PTE) will produce a voltage $V_{PTE}=(S_1-S_2)\Delta T_e$, where $S_{1,2}$ (in $V K^{-1}$) are the thermoelectric powers (or Seebeck coefficients) and $\Delta T_e$ is the HEs temperature difference in the two SLG regions\cite{gaborScience2011,koppensNN2014}. HEs in SLG can recombine radiatively to give broadband emission\cite{freitagNN2010,kimNN2015,luiPRL2010,chenNature2011,stoehrPRB2010,liuPRB2010} and the timescale/mechanism of the HEs relaxation has implications for the use of SLG in mode-locked lasers\cite{bonaccorsoNP2010,sunACSNano2010,FerrN2015}. SLG can be used to fabricate broadband and gate-tuneable optical frequency converters\cite{soaviNN2018,jiangNP2018,alexanderACSP2017,RomaNRM3}. However, in these devices the high $T_e$ ($\sim 10^3K$) induced by the optical excitation\cite{soaviNN2018,bridaNC2013,luiPRL2010} can significantly modify (e.g. by reducing the THG efficiency, THGE, defined as the ratio between the THG and incident intensities) the SLG nonlinear optical response\cite{soaviNN2018}.

Here we demonstrate that for THG in SLG the cubic dependence $I_{3\omega_0} \propto I_{\omega_0}^3$\cite{shen1984,boyd2003} fails when $T_e>>T_l$ is taken into account. We show that, more generally, THG follows a power-law $I_{3\omega_0} \propto I_{\omega_0}^x$, with the exponent $x$ dependent on $E_F$. This strong dependence of $T_e$, and thus of THGE, over both $E_F$ and $I_{\omega_0}$ has strong impact on the performance of nonlinear photonic devices based on SLG, such as optical switches and frequency converters.
\begin{figure}
\centerline{\includegraphics[width=90mm]{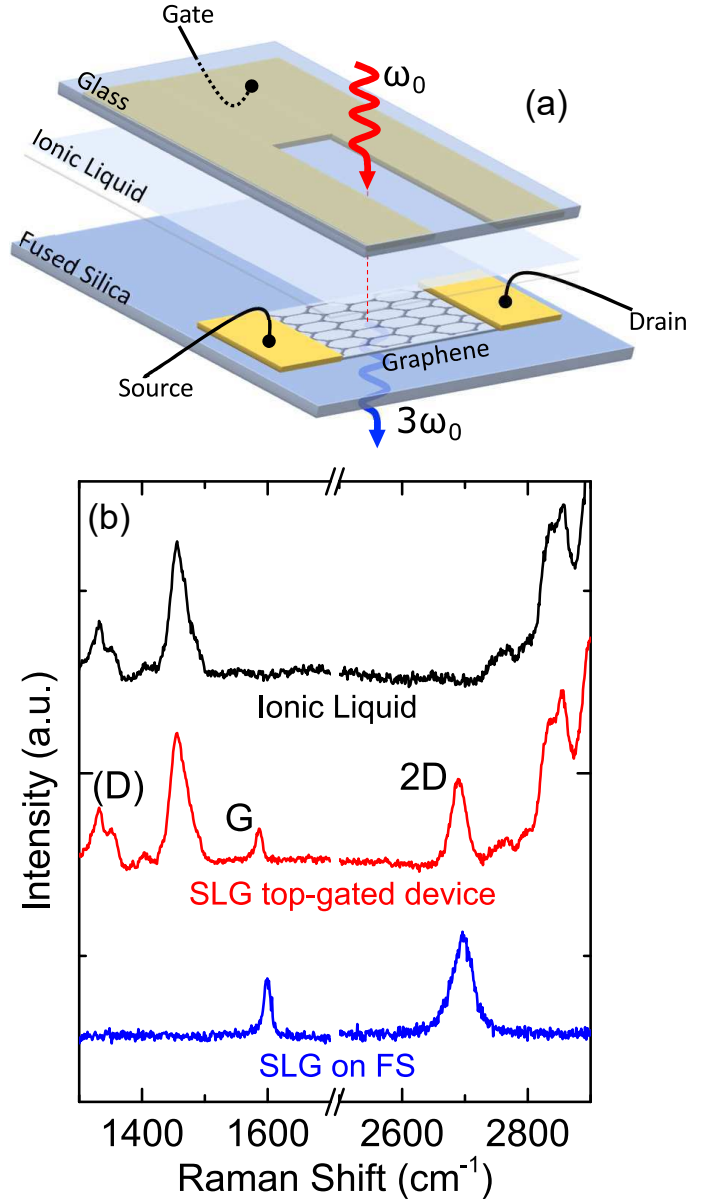}}
\caption{a) Schematic of THG device. $E_F$ tuning is obtained by IL top-gating. Measurements are performed in transmission. b) 514nm Raman spectra of SLG after transfer on FS (blue), SLG top-gated device (red) and IL (black)}
\label{fig:sample_raman}
\end{figure}

We use Chemical Vapor Deposition (CVD) SLG transferred on Fused Silica (FS) and gated by ionic liquid (IL), Fig.\ref{fig:sample_raman}a. SLG is grown on Cu (99.8$\%$ pure, 25$\mu$m thick), as for Ref.\cite{liScience2009}. This is then transferred on FS by polymer-assisted Cu wet etching\cite{bonaccorsoMatTod2012}, using polymethyl methacrylate (PMMA). SLG is characterized by Raman spectroscopy with a Renishaw inVia spectrometer. The 514nm Raman spectrum of SLG after transfer is shown in Fig.\ref{fig:sample_raman}b. The 2D peak is a single Lorentzian with full width at half maximum FWHM(2D)$\sim$36cm$^{-1}$, a signature of SLG\cite{ferrariPRL2006}. The position of the G peak, Pos(G), is$\sim$1599cm$^{-1}$, with FWHM(G)$\sim$13cm$^{-1}$. The 2D peak position is Pos(2D)$\sim$2696cm$^{-1}$, while the 2D to G peak intensity and area ratios, I(2D)/I(G) and A(2D)/A(G), are $\sim$1.7 and $\sim$4.67, indicating a p-doping$\sim$250-300meV\cite{dasNN2008,baskoPRB2009}. The absence of the D peak shows that there are no significant defects. In order to gate the SLG, we fabricate source and drain contacts by evaporating 7nm/70nm Cr/Au. Cr is used to improve Au adhesion. We etch the SLG outside the channel using an oxygen plasma. As gate electrode we use 7nm/70nm Cr/Au on a 1mm thick microscope slide. During evaporation, we cover part of the slide to have a transparent region$\sim$1cm$^2$ for optical measurements. We use 50$\mu$m double-sided tape as a spacer between gate electrode and SLG. We then align the SLG channel and the non-evaporated window on the gate electrode and place the IL, Diethylmethyl(2-methoxyethyl)ammoniumbis-(triflouromethylsulfonyl)imide ($C_6H_{20}F_6N_2O_5S_2$), between SLG and the gate electrode.
\begin{figure}
\centerline{\includegraphics[width=90mm]{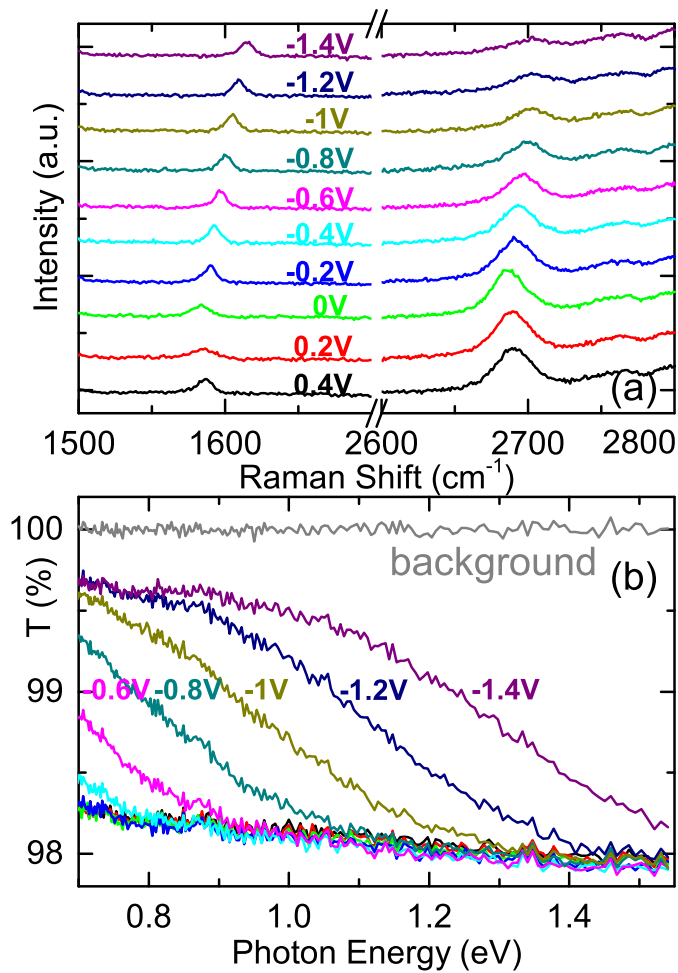}}
\caption{a) Raman; b) transmission spectra of SLG top-gated device at different $V_G$. The background (100$\%$) for the transmission spectra is defined as the transmission of the device without SLG.}
\label{fig:raman_abs}
\end{figure}
\begin{figure}
\centerline{\includegraphics[width=90mm]{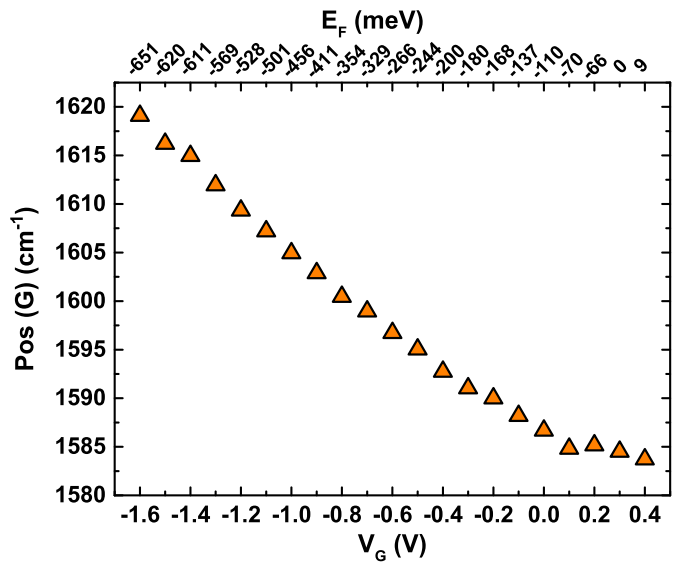}}
\caption{Pos(G) as a function of $V_G$ from the Raman measurements in Fig.\ref{fig:raman_abs}a. $E_F$ (top horizontal axis) is obtained as detailed in Ref.\cite{dasNN2008}.}
\label{fig:raman_Ef}
\end{figure}
\begin{figure}
\centerline{\includegraphics[width=90mm]{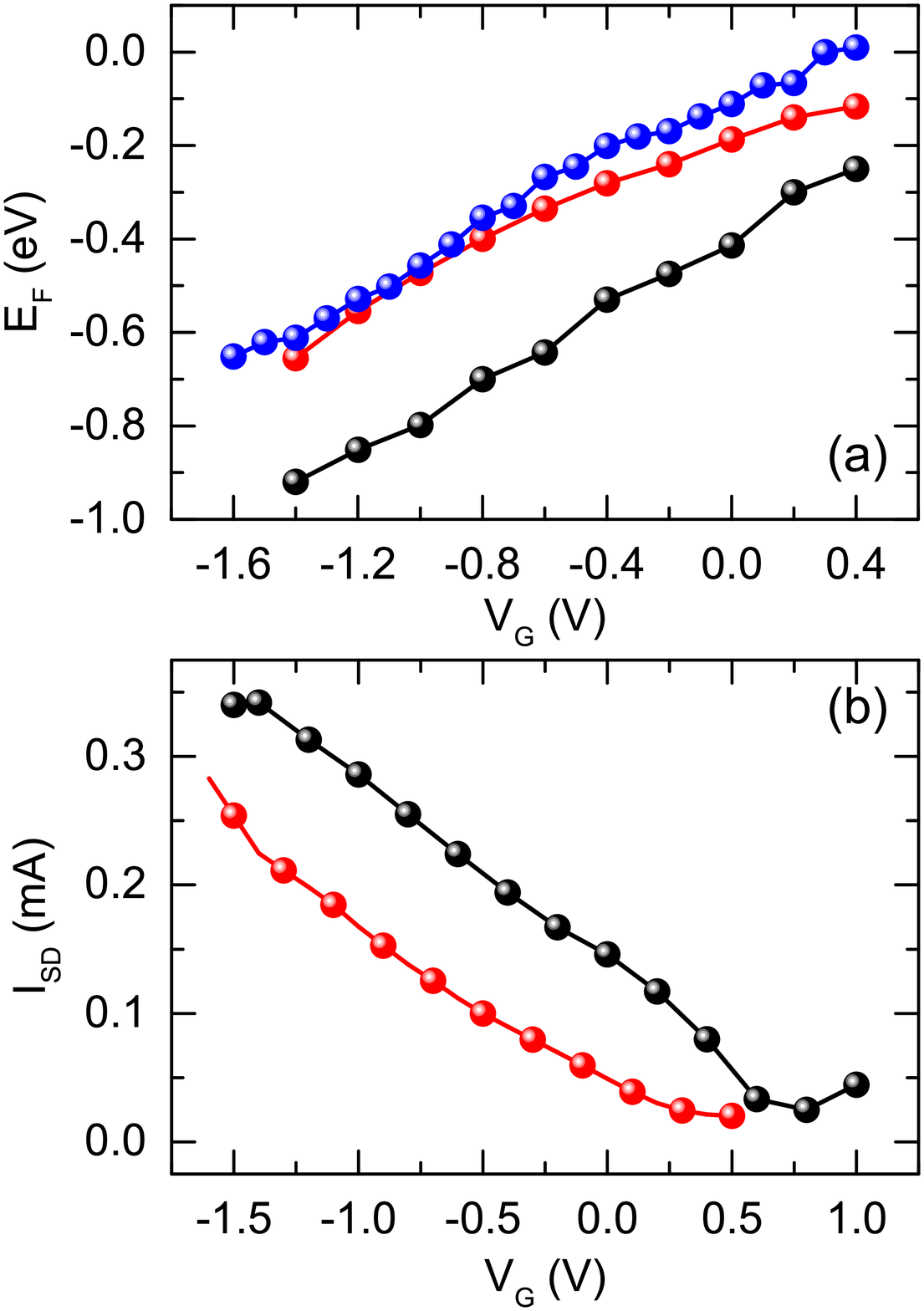}}
\caption{a) $E_F$ as a function of $V_G$ obtained from Raman analysis as in Fig.\ref{fig:raman_Ef} (blue dots) and from transmission measurements in Fig.\ref{fig:raman_abs}b (red dots). The black dots are $E_F$ during THG experiments calculated from $I_{SD}$. b) $I_{SD}$ as a function of $V_G$ before (red dots) and during (black dots) THG experiments.}
\label{fig:Ef_Isd}
\end{figure}

The 514nm Raman spectra of IL and of SLG at a gate voltage $V_G$=0V are shown in Fig.\ref{fig:sample_raman}b. For SLG, Pos(G) is$\sim$1587cm$^{-1}$, with FWHM(G)$\sim$14cm$^{-1}$. Pos(2D)$\sim$2691cm$^{-1}$, FWHM(2D)=32cm$^{-1}$, with I(2D)/I(G) and A(2D)/A(G)$\sim$2.9 and $\sim$5.9, respectively, indicating a p-doping$\sim$200meV\cite{dasNN2008}. Figs.\ref{fig:raman_abs}a,b plot the Raman and transmission spectra as a function of $V_G$ from 0.5 to -1.5V with steps of 0.1V for a source-drain voltage $V_{SD}$=0.2V. From the Raman spectra at different $V_G$ we estimate $E_F$. This is done by monitoring the evolution of Pos(G) as a function of $V_G$, as shown in Fig.\ref{fig:raman_Ef}\cite{dasNN2008,baskoPRB2009}. The relation between $E_F$ and $V_G$ can also be derived from the transmission measurements. For each $V_G$, we measure both transmission, Fig.\ref{fig:raman_abs}b, and source-drain current $I_{SD}$, Fig.\ref{fig:Ef_Isd}(b) (red circles). The transmission of the gated device never reaches 100$\%$, this being defined as the transmission of the device without SLG. This non-saturable residual absorption ($\alpha_{res}$) of SLG\cite{makSolidStateComms2012} originates from intra-band electronic transitions, enabled by disorder\cite{makSolidStateComms2012}. From Fig.\ref{fig:raman_abs}b we get $\alpha_{res} \sim 0.2-0.4\%$, by taking the difference between the background (grey curve) and the SLG transmission at 0.8eV for $V_G=-1.4V$. The transition from intra- to inter-band absorption, at $T_e=0$, occurs when the energy of the photons is $\hbar\omega_T=2E_F$. We thus estimate $\hbar\omega_T$ from the half-maximum of each transmission curve and calculate $E_F=\hbar\omega_T/2$, as in Fig.\ref{fig:Ef_Isd}(a) (red circles). This estimate is in good agreement with that derived from the Raman analysis (blue circles in Fig.\ref{fig:Ef_Isd}a).
\begin{figure}
\centerline{\includegraphics[width=86mm]{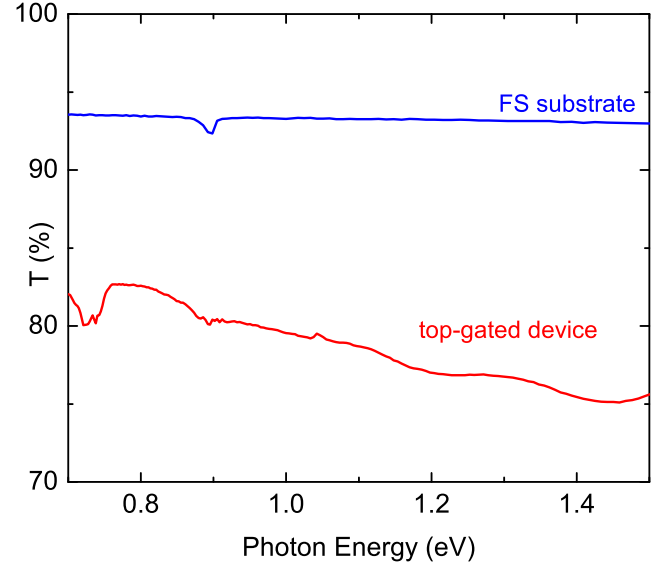}}
\caption{UV-VIS transmission curves for the device substrate (FS) and the final device (IL and FS) without SLG. The IL shows an absorption peak at $\sim$0.6eV.}
\label{fig:abs_substrate}
\end{figure}

THG measurements are then performed at room temperature (RT). We excite the sample with the idler beam of an Optical Parametric Oscillator (OPO, Coherent) at 0.69eV ($\sim$1.8$\mu$m) pumped by a mode-locked Ti:Sa laser (Coherent) with 150fs pulse duration, 80MHz repetition rate and 4W average power at 1.55eV. The OPO idler is focused by a 40X reflective objective (Ag coating, numerical aperture NA=0.5) to avoid chromatic aberrations. The THG signal is collimated by an 8mm lens and delivered to a spectrometer (Horiba iHR550) equipped with a nitrogen cooled Si charged-coupled-device (CCD). The idler spot-size is$\sim$4.7$\mu$m, the pulse duration$\sim$300fs and the polarization is linear. We use a Keithley 2612B dual channel Source Measure Unit both to apply $V_G$ and $V_{SD}$ and to read $I_{SD}$. $V_G$ is tuned between -1.5 and +0.5V while $V_{SD}$ is kept at 0.2V. For THG measurements we proceed we tune $V_G$ (10 points between -1.5 and +0.5V) and scan the power (7 points between 1 and 4 mW). The incident excitation power is estimated at the sample position by considering the losses of the objective, by measuring the power before and after the objective when the sample is removed. For each power (at a fixed $V_G$), we measure the THG signal by using 10s acquisitions and 3 accumulations. Thus, SLG is kept at a given $V_G$ for 210s before moving to next $V_G$. During THG experiments we also measure $I_{SD}$. By comparing the transconductance ($I_{SD}$ as a function of $V_G$) during the transmission, Fig.\ref{fig:Ef_Isd}(b) (red curve), and THG measurements, Fig.\ref{fig:Ef_Isd}(b) (black curve), we observe an increase in SLG doping. We thus estimate $E_F$ during THG experiments based on $I_{SD}$, Fig.\ref{fig:Ef_Isd}a (black curve). In order to estimate the emitted THG power, we take into consideration the losses of the system. The major ones are the absorption of the device without SLG (FS substrate and IL), the grating efficiency, and the CCD quantum efficiency. We also consider the CCD gain. The transmission of the FS substrate is$\sim$93$\%$, Fig.\ref{fig:abs_substrate}. The IL transmission is frequency dependent, Fig.\ref{fig:abs_substrate} (red curve). We use the spectrometer specs\cite{SymphonySpecs} to estimate losses due to grating and CCD efficiencies. We account for the$\sim$7 CCD gain, i.e. the number of electrons necessary for 1 count\cite{SymphonySpecs}.

The THG intensity $I_{3\omega_0}$ under normal incidence can be written as\cite{soaviNN2018}:
\begin{equation}\label{eq:THG_intensity}
I_{3\omega_0}=f(\omega_0)\frac{I^3_{\omega_0}}{4\epsilon^4_0 c^4}\left|\sigma^{(3)}_{\ell\ell\ell\ell}(\omega_0,E_F,T_e)\right |^2~,
\end{equation}
where $\epsilon_0\sim 8.85\times 10^{-12}{\rm C (V m)^{-1}}$ and $c=3\times10^8$m/s are the vacuum permittivity and the speed of light; $f(\omega_0)=n^{-3}_1(\omega_0)n_2(3\omega_0)[n_1(3\omega_0)+n_2(3\omega_0)]^{-2}$ in which $n_{i=1,2}(\omega)$ is the IL refractive index ($i=1$) and substrate ($i=2$). $\sigma^{(3)}_{\ell\ell\ell\ell}$ is the SLG third-order nonlinear optical conductivity tensor, calculated through a diagrammatic technique, with the light-matter interaction in the scalar potential gauge in order to capture all intra-, interband and mixed transitions\cite{rostamiPRB2016,soaviNN2018}. According to the C$_{\rm 6v}$ point group symmetry of SLG on a substrate, the relative angle between laser polarization and the SLG lattice is not important for the third-order response\cite{soaviNN2018}. Thus, we assume the incident polarization, $\hat{\ell}$, to lie along the zigzag direction of the lattice, $\hat{x}$, without loss of generality\cite{soaviNN2018}. For IL we use $n_1(\omega_0) \sim$1.44\cite{refindex} and for FS $n_2(3\omega_0)\sim$1.42\cite{refindex}. At first sight, Eq.(\ref{eq:THG_intensity}) predicts a cubic dependence $I_{3\omega_0} \propto I_{\omega_0}^3$. However, $I_{3\omega_0}$ is modulated also by $\sigma^{(3)}_{\ell\ell\ell\ell}$, which is a function of $\omega_0$, $E_F$ and $T_e$. The first two parameters, $\omega_0$ and $E_F$, can be controlled by tuning the excitation photon energy and by applying an external $V_G$. On the other hand, $T_e$ cannot be directly controlled by an external input, and its value is affected by the amount of energy that is transferred from light to the SLG electrons. T$_e$ can be calculated from the Boltzmann equation, taking into account the role of intra- and inter-band e-e scattering and the population of the optical phonon modes\cite{tomadinPRB2013}. An estimate can also be obtained with the following approach\cite{soaviNN2018}. When a pulse of duration $\Delta t$ and fluence ${\cal F}$ [$Jm^{-2}$] photoexcites SLG, an average power per unit area $P/A = (\alpha + \alpha_{res}) {\cal F} / \Delta t$ is absorbed by the electronic system, where $\alpha$ is the saturable SLG absorption, due to inter-band electronic transitions. $\alpha$ is a function of $\omega_0$, the chemical potentials in the conduction and valence bands ($\mu_c$ and $\mu_v$) and $T_e$. The variation $dU$ of the energy density in a time interval $dt$ is $dU=(P/A)dt$. The corresponding $T_e$ increase is $dT_e=dU/c_v$, where $c_v$ is the electronic heat capacity of the photoexcited SLG. When the pulse is off, $T_e$ relaxes towards T$_0$ on a time-scale $\tau$. This reduces $T_e$ by $dT_e=-(T_e/\tau)dt$ in a time interval $dt$. Thus\cite{soaviNN2018}:
\begin{equation}\label{eq:rate_eq_temp}
\frac{dT_{\rm e}}{dt}=\frac{\alpha + \alpha_{res}}{c_{v}}\frac{\cal F}{\Delta t}-\frac{T_{\rm e}-T_0}{\tau}~,
\end{equation}
\begin{figure}
\centerline{\includegraphics[width=90mm]{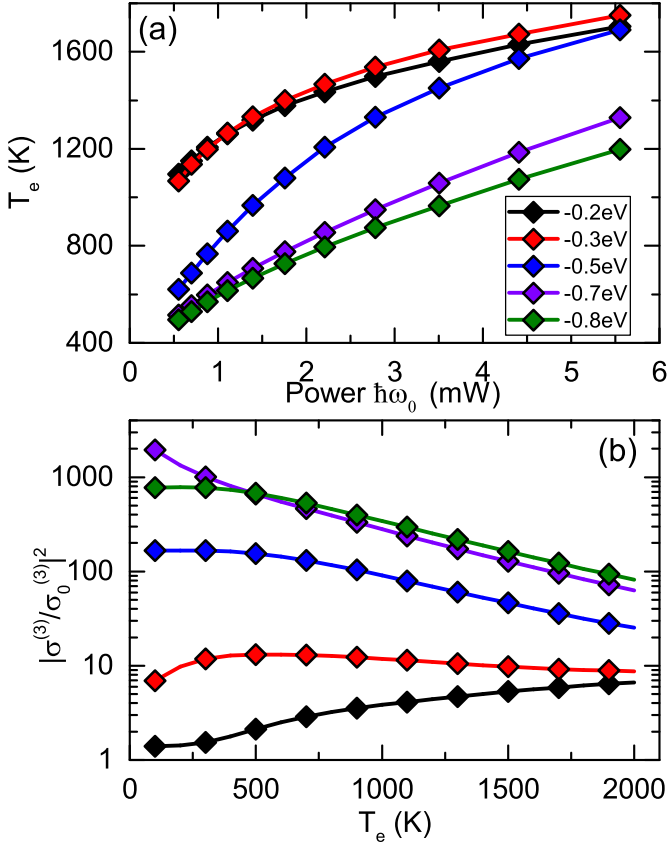}}
\caption{a) $T_e$ from Eq.(\ref{eq:steady_temp}) as a function of the incident power for $\hbar\omega_0$=0.69eV and different $E_F$. b) $|\sigma^{(3)}/\sigma_0^{(3)}|^2$ as a function of $T_e$  for $\hbar\omega_0$=0.69eV and different $E_F$ in a single-chemical potential model}
\label{fig:Te_sigma}
\end{figure}
If the pulse duration is: (i) much longer than$\sim$20fs, i.e. the time-scale for the e distribution to relax to the Fermi-Dirac profile in both bands\cite{bridaNC2013,breusingPRB2011}; (ii) comparable to the time-scale$\tau\sim100-200fs$ needed to heat the optical ph modes\cite{bridaNC2013,breusingPRB2011,lazzeriPRL2005}, the electronic system reaches a steady-state during the pulse, with T$_e$ obtained from Eq.(\ref{eq:rate_eq_temp}):
\begin{equation}\label{eq:steady_temp}
T_e= T_0+\tau\frac{\alpha+\alpha_{res}}{c_v}\frac{\cal F}{\Delta t}~.
\end{equation}
Fig.\ref{fig:Te_sigma}(a) plots $T_e$ from Eq.(\ref{eq:steady_temp}) for our experimental conditions: excitation power$\sim$0.5mW to 5mW, $E_F \sim$-0.8 to -0.2eV, $\hbar\omega_0$=0.69eV, $T_0$=300K, $\tau$=100fs, $\alpha_{res}=0.4\%$ and $\Delta t$=300fs. An increase of excitation power induces an increase of $T_e$, thus a modulation of $\sigma^{(3)}$. The increase in $T_e$ is also modulated by changes in $E_F$, as this affects $\alpha$ of SLG (Fig.\ref{fig:raman_abs}b). Fig.\ref{fig:Te_sigma} shows the $T_e$ dependence of $\sigma^{(3)}$ in the 0-2000K range and for different $E_F$. Fig.\ref{fig:Te_sigma}b plots $|\sigma^{(3)}/\sigma_0^{(3)}|^2$, with $\sigma^{(3)}_0 = N_f e^4 \hbar v_F^2/[32 \pi (1{\rm eV})^4]$\cite{soaviNN2018}. The quantity $N_f=4$ is the number of fermion flavors in SLG and $v_F \sim 10^6$m/s is the Fermi velocity, thus $\sigma^{(3)}_0 \sim 4.2 \times 10^{-24} A m^2/V^3$\cite{soaviNN2018}. Fig.\ref{fig:Te_sigma}b shows that, depending on $E_F$, $\sigma^{(3)}$ will either increase (e.g. $E_F$=-0.2eV in Fig.\ref{fig:Te_sigma}b) or decrease (e.g. $E_F$=-0.5 to -0.8eV in Fig.\ref{fig:Te_sigma}b) with increasing $T_e$. This results in a deviation from the cubic dependence $I_{3\omega_0} \propto I_{\omega_0}^3$.
\begin{figure}
\centerline{\includegraphics[width=90mm]{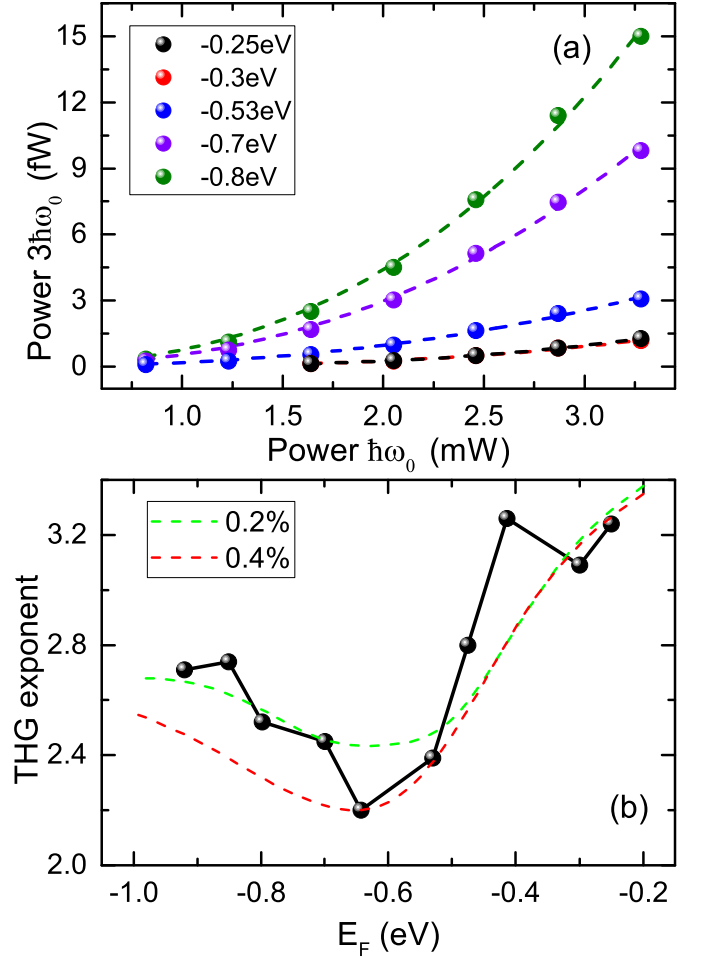}}
\caption{a) $3\hbar\omega_0$ power as a function of incident power for different $E_F$. The dotted lines are obtained from the power-law $y=a \cdot x^b$, with a and b fitting parameters. b) THG exponent from fitting ($y=a \cdot x^b$) the power-dependent THG (black points) and from theory (dotted-lines) for different $\alpha_{res}$ and $\hbar\omega_0=0.69eV$.}
\label{fig:exp}
\end{figure}

Fig.\ref{fig:exp}(a) plots the experimental THG power dependence for $\hbar\omega_0$=0.69eV. For the same values of incident power we do not detect any THG signal from FS/IL, i.e. outside the area covered by SLG. For a fixed incident power, the THG power increases as we go to more negative values of $E_F$. This $E_F$ dependent enhancement of the THG signal arises from logarithmic resonances in the imaginary part of the nonlinear conductivity of SLG due to resonant multiphoton transitions\cite{soaviNN2018}. As seen in Fig.\ref{fig:Te_sigma}(b), this leads to a non-monotonic dependence of the nonlinear conductivity on T$_e$ for different $E_F$. We fit the experimental data relative to our THG power-dependent measurements (circles in Fig.\ref{fig:exp}a) with the power law $y=a \cdot x^b$ (dotted lines in Fig.\ref{fig:exp}a), where $y$ is the $3\hbar\omega_0$ power, $x$ is the incident power and $a$, $b$ are fitting parameters. Fig.\ref{fig:exp}a shows that the power-law approximation gives excellent fits to the data, if we allow $b$ to depend on $E_F$. Fig.\ref{fig:exp}b plots $b$ (i.e. the THG exponent) from this fit (black circles) as a function of $E_F$. The dotted lines in Fig.\ref{fig:exp}b are the theoretical $b$ (THG exponent) calculated as follows: (i) $T_e$ and corresponding chemical potentials in conduction and valence bands as a function of incident power are derived from Eq.(\ref{eq:steady_temp}), for $\hbar\omega_0=0.69eV$ and different $E_F$; (ii) we use these to calculate $\sigma^{(3)}$ as a function of incident power. To this end, we first calculate the $T_e=0$ expression of the third-order nonlinear conductivity\cite{rostamiPRB2016} and then utilize the response function in Ref.\cite{tomadinPRB2013}, to express the conductivity at finite T as a weighted integral over $E_F$ of the SLG conductivity at $T_e=0$; (iii) we substitute the calculated $\sigma^{(3)}$ into Eq.(\ref{eq:THG_intensity}) to obtain the theoretical THG intensity; (iv) we fit the THG intensity with $y=a \cdot x^b$. For the estimate of $T_e$ we use $\alpha_{res}$=0.2\% and 0.4$\%$, as derived from Fig.\ref{fig:raman_abs}b. We find that the THG exponent varies between$\sim$2 and 3.4, with a non-monotonic dependence on $E_F$ and a minimum at $E_F\sim$0.6eV for $\hbar\omega_0=0.69eV$. An increase of the incident power affects $T_e$ and $\sigma^{(3)}$. This induces deviations from the cubic power law. To the best of our knowledge, this non-cubic behavior of the THG signal was not reported before in SLG or any other material. Most experiments on SLG and layered materials took the observation of a cubic power law as a proof of THG\cite{kumarPRB2013,hongPRX2013,wangACS2014,woodward2Dmat2017}. In SLG, this cubic dependence was also used to calculate $\rchi^{(3)}$\cite{kumarPRB2013,hongPRX2013,woodward2Dmat2017}. This approach has two limitations: 1) the nonlinear susceptibilities are well defined only in three-dimensional materials, since they involve a polarization per unit volume\cite{soaviNN2018}, thus $\rchi^{(3)}$ should not be used for SLG; 2) a power-law fit of THG in SLG must take into account $E_F$ and $T_e$ under the specific experimental conditions. In other words, $\rchi^{(3)}$ in SLG must be calculated as a function of both $E_F$ and $T_e$.

In summary, hot electrons strongly affect the third-order nonlinear optical response of single-layer graphene and alter the cubic dependence of the third harmonic generation signal and its efficiency. Upon ultrafast ($\sim$100fs) excitation, $T_e$ in single-layer graphene can be as high as 10$^3$K also when $E_F>2\hbar\omega_0$, due to thermal broadening of the Fermi-Dirac distribution and residual absorption $\alpha_{res}$. Thus, $T_e$ is affected by both $E_F$ and $I_{\omega_0}$. Changes in $T_e$ can modify the third harmonic generation efficiency and thus the performances of nonlinear photonic and optoelectronic devices, such as optical switches and frequency converters.

We acknowledge funding from EU Graphene Flagship, ERC Grant Hetero2D, EPSRC Grants EP/K01711X/1, EP/K017144/1, EP/N010345/1, EP/L016087/1 and the Swedish Research Council (VR 2018-04252).


\begin{thebibliography}{100}
\bibitem{kittel1996} C. Kittel, \textit{Introduction to Solid State Physics} (Wiley New York, New York, 1996).

\bibitem{farm1992} S. Farm, R. Storz, H. K. Tom, and J. Bokor, Phys. Rev. Lett. \textbf{68}, 2834 (1992).

\bibitem{dellavalle2012} G. Della Valle, M. Conforti, S. Longhi, G. Cerullo, and D. Brida, Phys. Rev. B \textbf{86}, 155139 (2012).

\bibitem{lazzeriPRL2005} M. Lazzeri, S. Piscanec, F. Mauri, A. C. Ferrari, and J. Robertson, Phys. Rev. Lett. \textbf{95}, 236802 (2005).

\bibitem{bridaNC2013} D. Brida, A. Tomadin, C. Manzoni, Y. J. Kim, A. Lombardo, S. Milana, R. R. Nair, K. S. Novoselov, A. C. Ferrari, G. Cerullo, M. Polini, Nature Commun. \textbf{4}, 1987 (2013).

\bibitem{tomadinPRB2013} A. Tomadin, D. Brida, G. Cerullo, A. C. Ferrari, and M. Polini, Phys. Rev. B \textbf{88}, 035430 (2013).

\bibitem{shank1983} C. V. Shank, R. Yen, C. Hirlimann, Phys. Rev. Lett.\textbf{50}, 454 (1983).

\bibitem{schoenlein1987} R. W. Schoenlein, W. Z. Lin, J. G. Fujimoto, and G. L. Eesley, Phys. Rev. Lett. \textbf{58}, 1680 (1987).

\bibitem{vallee2001} F. Vall\'ee Acad. Sci. Paris~\textbf{2}, 1469 (2001).

\bibitem{hohlChemPhys2000} J. Hohlfeld, S.-S. Wellershoff, J. G\"udde, U. Conrad, V. J\"ahnke, and E. Matthias, Chem. Phys. \textbf{251}, 237 (2000).

\bibitem{shah2013} J. Shah, \textit{Ultrafast Spectroscopy of Semiconductors and Semiconductor Nanostructures} (Springer Science \& Business Media, New York, 2013).

\bibitem{soaviAOM2016} G. Soavi, F. Scotognella, G. Lanzani, and G. Cerullo, Adv. Opt. Mat. \textbf{4}, 1670 (2016).

\bibitem{weiNanop2017} L. Wei and J. G. Valentine, Nanophot.\textbf{6}, 177 (2017).

\bibitem{mukhNanoLett2013} S. Mukherjee, F. Libisch, N. Large, O. Neumann, L. V. Brown, J. Cheng, J. B. Lassiter, E. A. Carter, P. Nordlander, N. J. Halas, Nano Lett. \textbf{13}, 240 (2013).

\bibitem{rossJAP1982} R. T. Ross, A. J. Nozik, J. Appl. Phys. \textbf{53}, 3813 (1982).

\bibitem{tisdScience2010} W. A. Tisdale, K. J. Williams, B. A. Timp, D. J. Norris, E. S. Aydil, X. Y. Zhu, Science \textbf{328}, 1543 (2010).

\bibitem{stiensProcSPIE2006} J. Stiens, G. Shkerdin, V. Kotov, W. Vandermeiren, C. D. Tandt, G. Borghs, R. Vounckx, Proc. SPIE \textbf{6189}, 61890Y (2006).

\bibitem{shephProcIEEE1970} F. D. Shepherd, A. C. Yang, R. W. Taylor, Proc. IEEE \textbf{58}, 1160 (1970).

\bibitem{frankPRL1961} P. A. Franken, A. E. Hill, C. W. Peters, G. Weinreich, Phys. Rev. Lett. \textbf{7}, 118 (1961).

\bibitem{terhPRL1962} R. W. Terhune, P. D. Maker, C. M. Savage, Phys. Rev. Lett. \textbf{8}, 404 (1962)

\bibitem{shen1984} Y. R. Shen, \textit{The Principles of Nonlinear Optics} (John Wiley \& Son, New York, 1984).

\bibitem{boyd2003} R. W. Boyd, \textit{Nonlinear Optics} (Academic Press, New York, 2003).

\bibitem{kumarPRB2013} N. Kumar, J. Kumar, C. Gerstenkorn, R. Wang, H.-Y. Chiu, A. L. Smirl, and H. Zhao, Phys. Rev. B \textbf{87}, 121406 (2013).

\bibitem{burnsPRB1971} W. K. Burns, N. Bloembergen, Phys. Rev. B \textbf{4}, 3437 (1971).

\bibitem{hohlApplPhysA1995} J. Hohlfeld, D. Grosenick, U. Conrad, E. Matthias, Appl. Phys. A \textbf{60}, 137 (1995).

\bibitem{guoPRL2001} C. Guo, G. Rodriguez, A. J. Taylor, Phys. Rev. Lett. \textbf{86}, 1638 (2001).

\bibitem{hohlAPB1996} J. Hohlfeld, U. Conrad, E. Matthias, Appl. Phys. B \textbf{63}, 541 (1996).

\bibitem{mooreOptLett1999} K. L. Moore, T. D. Donnelly, Opt. Lett. \textbf{24}, 990 (1999).

\bibitem{papaOptComm1997} N. Papadogiannis, S. Moustaizis, Opt. Commun. \textbf{137}, 174 (1997).

\bibitem{tomPRL1988} H. W. K. Tom, G. D. Aumiller, C. H. Brito-Cruz, Phys. Rev. Lett. \textbf{60}, 1438 (1988).

\bibitem{saePRL1991} P. Saeta, J. K. Wang, Y. Siegal, N. Bloembergen, E. Mazur, Phys. Rev. Lett. \textbf{67}, 1023 (1991).

\bibitem{bistritzerPRL2009} R. Bistritzer, A. H. MacDonald, Phys. Rev. Lett. \textbf{102}, 206410 (2009).

\bibitem{betzPRL2012} A. C. Betz, F. Vialla, D. Brunel, C. Voisin, M. Picher, A. Cavanna, A. Madouri, G. F\'eve, J. M. Berroir et al., Phys. Rev. Lett. \textbf{109}, 056805 (2012).

\bibitem{tielrooijNP2013} K.J. Tielrooij, J. Song, S. A. Jensen, A. Centeno, A. Pesquera, A. Z. Elorza, M. Bonn, L. Levitov, F. Koppens, Nature Phys. \textbf{9}, 248 (2013).

\bibitem{mikhailovPRB2016} S. A. Mikhailov, Phys. Rev. B \textbf{93}, 085403 (2016).

\bibitem{rostamiPRB2016} H. Rostami, M. Polini, Phys. Rev. B \textbf{93}, 161411 (2016).

\bibitem{chengNJP2014} J. L. Cheng, N. Vermeulen, J. E. Sipe, New J. Phys. \textbf{16}, 053014 (2014).

\bibitem{lazzeriPRB2006} M. Lazzeri, S. Piscanec, F. Mauri, A. C. Ferrari, J. Robertson, Phys. Rev. B \textbf{73}, 155426 (2006).

\bibitem{bonaccorsoNP2010} F. Bonaccorso, Z. Sun, T. Hasan, A. C. Ferrari, Nature Photon. \textbf{4}, 611 (2010).

\bibitem{RomaNRM3} M. Romagnoli, V. Sorianello, M. Midrio, F.H.L. Koppens, C. Huyghebaert, D. Neumaier, P. Galli, W. Templ, A. D'Errico, A.C. Ferrari, Nature Reviews Materials, \textbf{3}, 392-414, (2018)

\bibitem{FerrN2015} A. C. Ferrari, F. Bonaccorso, V. Fal'ko, K. S. Novoselov, S. Roche, P. Boggild, S. Borini, F. H. L. Koppens, V. Palermo, N. Pugno, \emph{et al.}, Nanoscale \textbf {7}, 4598 (2015).

\bibitem{gaborScience2011} N. M. Gabor, J. C. W. Song, Q. Ma, N. L. Nair, T. Taychatanapat, K. Watanabe, T. Taniguchi, L. S. Levitov, P. Jarillo-Herrero, Science \textbf{334}, 648 (2011).

\bibitem{koppensNN2014} F. H. L. Koppens, T. Mueller, P. Avouris, A. Ferrari, M. Vitiello, M. Polini, Nature Nanotech. \textbf{9}, 780 (2014).

\bibitem{freitagNN2010} M. Freitag, H.Y. Chiu, M. Steiner, V. Perebeinos, P. Avouris, Nature Nanotech. \textbf{5}, 497 (2010).

\bibitem{kimNN2015} Y. D. Kim, H. Kim, Y. Cho, J. H. Ryoo, C.-H. Park, P. Kim, Y. S. Kim, S. Lee, Y. Li et al. Nature Nanotech. \textbf{10}, 676 (2015).

\bibitem{luiPRL2010} C. H. Lui, K. F. Mak, J. Shan, and T. F. Heinz, Phys. Rev. Lett. \textbf{105}, 127404 (2010).

\bibitem{chenNature2011} C. F. Chen, C. H. Park, B. W. Boudouris, J. Horng, B. Geng, C. Girit, A. Zettl, M. F. Crommie, R. A. Segalman, S. G. Louie, F. Wang, Nature \textbf{471}, 617 (2011).

\bibitem{stoehrPRB2010} R. J. Stoehr, R. Kolesov, J. Pflaum, J. Wrachtrup, Phys. Rev. B \textbf{82}, 121408(R) (2010).

\bibitem{liuPRB2010} W.-T. Liu, W. Wu, P. J. Schuck, M. Salmeron, Y. R. Shen, F. Wang, Phys. Rev. B \textbf{82}, 081408(R) (2010).

\bibitem{sunACSNano2010} Z. Sun, T. Hasan, F. Torrisi, D. Popa, G. Privitera, F. Wang, F. Bonaccorso, D. M. Basko, A. C. Ferrari, ACS Nano \textbf{4}, 803 (2010).

\bibitem{soaviNN2018} G. Soavi, G. Wang, H. Rostami, D. G. Purdie, D. De Fazio, T. Ma, B. Luo, J. Wang, A. K. Ott et al., Nature Nanotech. \textbf{13}, 583 (2018).

\bibitem{jiangNP2018} T. Jiang, D. Huang, J. Cheng, X. Fan, Z. Zhang, Y. Shan, Y. Yi, Y. Dai, L. Shi et al., Nature Phot. \textbf{12}, 430 (2018).

\bibitem{alexanderACSP2017} K. Alexander, N. A. Savostianova, S. A. Mikhailov, B. Kuyken, D. Van Thourhout, ACS Phot. \textbf{4}, 3039 (2017).

\bibitem{liScience2009} X. Li, W. Cai, J. An, S. Kim, J. Nah, D. Yang, R. Piner, A. Velamakanni, I. Jung et al., Science \textbf{324}, 1312 (2009).

\bibitem{bonaccorsoMatTod2012} F. Bonaccorso, A. Lombardo, T. Hasan, Z. Sun, L. Colombo, A. C. Ferrari, Materials Today \textbf{15}, 564 (2012).

\bibitem{ferrariPRL2006} A. C. Ferrari, J. C. Meyer, V. Scardaci, C. Casiraghi, M. Lazzeri, F. Mauri, S. Piscanec, D. Jiang, K. S. Novoselov et al., Phys. Rev. Lett. \textbf{97}, 187401 (2006).

\bibitem{dasNN2008} A. Das, S. Pisana, B. Chakraborty, S. Piscanec, S. K. Saha, U. V. Waghmare, K. S. Novoselov, H. R. Krishnamurthy, A. K. Geim et al., Nature Nanotech. \textbf{3}, 210 (2008).

\bibitem{baskoPRB2009} D. M. Basko, S. Piscanec, A. C. Ferrari, Phys. Rev. B \textbf{80}, 165413 (2009).

\bibitem{makSolidStateComms2012} K. F. Mak, L. Ju, F. Wang, T. F. Heinz, Solid State Comms \textbf{152}, 1341 (2012).

\bibitem{SymphonySpecs} Symphony II 1024 x 256 Cryogenic Open-Electrode CCD Detector Quantum Efficiency and grating (300 gr/mm, Blazed 600 nm 510 19 140) relative efficiency from www.horiba.com.

\bibitem{refindex} https://www.sigmaaldrich.com.

\bibitem{breusingPRB2011} M. Breusing, S. Kuehn, T. Winzer, E. Mali\'c, F. Milde, N. Severin, J. P. Rabe, C. Ropers, A. Knorr, T. Elsaesser, Phys. Rev. B \textbf{83}, 153410 (2011).

\bibitem{wangACS2014} R. Wang, H. C. Chien, J. Kumar, N. Kumar, H.-Y. Chiu, H. Zhao, ACS Appl. Mat. Int. \textbf{6}, 314 (2014).

\bibitem{hongPRX2013} S. Y. Hong, J. I. Dadap, N. Petrone, P. C. Yeh, J. Hone, R. M. Osgood, Phys. Rev. X \textbf{3}, 021014 (2013).

\bibitem{woodward2Dmat2017} R. I. Woodward, R. T. Murray, C. F. Phelan, R. E. P. de Oliveira, T. H. Runcorn, E. J. R. Kelleher, S. Li, E. C. de Oliveira, G. J. M. Fechine et al. 2D Materials \textbf{4}, 011006 (2017).

\end{thebibliography}
\end{document}